\documentstyle[aps,preprint]{revtex}

\begin{document}
\draft
\tightenlines
\preprint{SNUTP-97-172, UOSTP-98-101}
\title{Fractal Diffraction Grating}
\author{Dongsu Bak$^a$, Sang Pyo Kim$^b$, Sung Ku Kim$^c$, 
Kwang-Sup Soh$^d$,
and Jae Hyung Yee$^e$}

\address{$^a$ Department of Physics, University of Seoul, Seoul 
130-743, Korea \\
$^b$ Department of Physics, Kunsan National University, 
Kunsan 573-701, Korea \\
$^c$ Department of Physics, Ewha Womans University, 
Seoul 120-750, Korea\\
$^d$ Department of Physics Education, Seoul National 
University, Seoul 151-742, Korea\\
$^e$ Department of Physics, Yonsei University, 
Seoul 120-749, Korea}
\date{\today}
\maketitle
\widetext
\begin{abstract}
We consider an optical diffraction grating  in which the spatial 
distribution of 
 open slits forms  a fractal set. 
The Fraunhofer diffraction patterns  
through the fractal  grating are   obtained analytically 
for 
the simplest triad Cantor type and its generalized version. The  
resulting interference patterns 
exhibit characteristics of  the original fractals and 
their scaling properties.
\end{abstract}

%\vskip .2cm
%\pacs{
\  \ \ \ \ \ \ \ \ (47.53.+n, 04.30.Nk, 01.55.+b)
%}

\section{Introduction}

 A widespread  interest in  fractal  geometry has  been 
generated  by  Mandelbrot with  his 
monumental work \cite{mandelbrot} that 
deals  with the geometry of  phenomena observed in 
many  fields   of science.   The  fractal   geometry is   
a  fast   growing subject   and  a 
cross-disciplinary  field  ranging  from  social   
science to   biological science   and  physics 
\cite{feder,krim,krug}.

In this  article we   consider another aspect   of fractal geometry,  
namely,  the Fraunhofer 
diffraction of an optical  grating whose slit distribution  is 
fractal. In the  ordinary optical 
gratings the slits are evenly distributed  forming  a one-dimensional 
lattice \cite{wolf}. We take,
instead, a 
fractal grating and study the  optical interference pattern 
%through it.
%formed 
by the light wave coming from the grating.  As a simple 
example  a 
Cantor set  type  grating is   considered, and analytic   
results are  obtained. The  resulting 
interference patterns show self-similar  structure and 
the same scaling  property 
 as those  in the 
fractal grating   itself. In  other words   
the Fraunhofer  diffraction  transforms the   fractal 
geometry of the optical grating to the fractal interference pattern.

In Sec. II the  simplest triad Cantor set  type grating is 
introduced,  and the interference 
amplitude is explicitly  evaluated. The  amplitude turns  
out to  be a  repeated product  of a 
function of a rescaled argument by $3^m$ for each integer $m$, 
which respects the scaling property of the 
diffraction grating.
%obtained from  a  scale  
%transformation of  factor  $3$ for  each repetition. 
In  Sec. III  we 
generalize the triad Cantor set to 
a $(2H+1)$-piece Cantor set, and similar results are 
obtained 
but with a different scale factor $2H+1$. 
In Sec. IV we  discuss  further generalizations 
of the model.
The ordinary diffraction grating has been an 
important tool  in optical spectroscopy. Although 
we have not  yet proposed any  practical application, 
this  kind of  fractal diffraction grating 
may be useful for some purposes.

\section{Cantor Diffraction Grating}
One of the simplest example 
of fractal sets is the Cantor set defined on
 a closed interval $[0,1]$. The first stage of construction 
consists of  dividing
the interval $[0,1]$ into three equal pieces, 
then removing the middle open 
interval, designated $(\frac{1}{3},
\frac{2}{3})$. At the second stage one divides each remaining
pieces into three and again remove each middle piece. Repeating 
this process infinitely many times one obtains a fractal set  
whose fractal dimension is 
$ D = \frac{\ln 2}{\ln 3}$\cite{mandelbrot,feder}.

Let us consider a diffraction grating whose slits are distributed 
as the Cantor set described above.
For convenience, we take the starting total transparent 
region to be $[-a, a]$. At the first 
stage we make the middle third 
$( - \frac{a}{3}, \frac{a}{3})$ opaque, in 
the second stage we again make the middle pieces of each of the 
two remaining thirds
opaque, and  so on   to $N$ times.  The  resulting grating  
consists of   fractally distributed 
$2^N$-slits of width
$\frac{2a}{3^N}$. Let this be called the Cantor diffraction grating of
$N$-th generation. Figure 1 illustrates such a construction for 
the case of $N =3$. 
The open slits are of equal widths but unevenly distributed. 
The coordinates of the
middle, beginning, and end points of the $j$-th 
slit are denoted by $x_j, x_j^-, x_j^+$,
respectively.

The Fraunhofer interference pattern of light through a 
Cantor diffraction grating
of $N$-th generation can be computed as
\begin{eqnarray}
\label{e01}
A_N &=& \int^{\infty}_{- \infty}dx  e^{ i kx \sin \theta} G(x)
\nonumber\\
&=& \sum_{j = 1}^{2^N} \frac{e^{i kx \sin \theta}}{i k \sin \theta} 
\vert^{x_j^+}_{x_j^-},
\end{eqnarray}
where $\theta$ is the  angle of light propagation with
respect to the normal vector of the grating plane, and $k$ is the wave 
vector. 
The grating function $G(x)$ is defined as
\begin{equation}
G(x)= \left\{ \begin{array}{ll}
              0, & x \in {\rm opaque~ region}, \\
              1, & x \in {\rm open~ region}.
              \end{array}  
      \right.
\end{equation}
Using the fact that the slit width $(x_j^{+} - x_j^{-})$ 
is $\frac{2a}{3^N}$,
we can write $A_N$ as
\begin{equation}
\label{e02}
A_N = \frac{2 \sin \Bigl( \frac{  k a \sin \theta }{3^N} \Bigr)}{k \sin
\theta} \sum_{j = 1}^{2^N} e^{ik  x_j \sin \theta}\,.
\end{equation}
The amplitude $A_N$ is factorized into two factors.  The first factor,
$2 \frac{\sin \Bigl(\frac{ k a \sin \theta }{3^N} \Bigr)}{ k \sin \theta}$,
is just the Fraunhofer diffraction amplitude of the single slit of
width $\frac{2a}{3^N}$. The second factor represents the interference of
the $2^N$-Cantor slits, which are expected to show the fractal nature
of the slit distribution.

In order to evaluate the summation in the second factor we first 
notice that
the slit-coordinates $x_j$ (the midpoint of the j-th open interval) 
can be represented by tertiary decimal numbers 
where only $\pm 1$ among $(-1, 0 , +1)$ appears as numerator. 
Explicitly, they are
\begin{eqnarray}
\label{e03}
x_1 &=& 2a \Bigl( \frac{- 1}{3} + \frac{ -1}{3^2} + 
\cdots + \frac{ -1}{3^{N -1}}
+ \frac{-1}{3^N} \Bigr),
\nonumber\\
x_2 &=& 2a \Bigl( \frac{- 1}{3} + \frac{ -1}{3^2} + 
\cdots + \frac{ -1}{3^{N -1}}
+ \frac{+1}{3^N} \Bigr),
\nonumber\\
x_3 &=& 2a \Bigl( \frac{- 1}{3} + \frac{ -1}{3^2} + 
\cdots + \frac{ +1}{3^{N -1}}
+ \frac{-1}{3^N} \Bigr), \nonumber\\
 \vdots &&~~~~~~~~~~~~~~~~~~ \vdots
\nonumber\\
x_{2^N} &=& 2a \Bigl( \frac{+ 1}{3} + \frac{ +1}{3^2} +
 \cdots + \frac{ +1}{3^{N -1}}
+ \frac{+1}{3^N} \Bigr) .
\end{eqnarray}
Using these coordinates we can evaluate the summation as
\begin{equation}
\label{e04}
\sum_{j = 1}^{ 2^N} e^{ ikx_j \sin \theta} = \prod_{ m=1}^{N}
2 \cos \Bigl(\frac{2 ka \sin \theta}{3^m} \Bigr).
\end{equation}
Substituting (\ref{e04}) to (\ref{e02}) we obtain 
the amplitude of the Cantor diffraction grating
as
\begin{eqnarray}
A_N &=& \Bigl(\frac{2}{3} \Bigr)^N 2a \frac{ \sin \Bigl(
\frac{ka \sin \theta}{3^N} \Bigr)}{\frac{ka \sin \theta}{ 3^N}}
\prod_{m = 1}^{N} \cos \Bigl (\frac{2 ka \sin \theta}{3^m} \Bigr)
\nonumber\\
&=& \Bigl(\frac{2}{3} \Bigr)^N 2a S_N ( \beta) F_N ( \beta),
\end{eqnarray}
where $\beta \equiv 2ka \sin \theta$.
Here $\Bigl(\frac{2}{3} \Bigr)^N 2a$ signifies that the peak amplitude
diminishes by the factor $\Bigl(\frac{2}{3} \Bigr)^N$ as the open length 
of the grating decreases by
the same factor. The $S_N (\beta)$ is the Fraunhofer diffraction amplitude of
the single slit of the width $\frac{2a}{3^N}$, and the $F_N ( \beta)$
codifies the structure of the $N$-th generation cantor set.

The repeated product of cosine functions 
in the $F_N (\theta)$ has a recursive 
nature as
\begin{equation}
F_{N+1} (\beta) = \cos \Bigl(\frac{\beta}{3} \Bigr) F_N \Bigl( 
\frac{\beta}{3} \Bigr),
\end{equation}
 In the limit $ N \rightarrow \infty$
we have
\begin{equation}
F_{\infty} (\beta) = \cos \Bigl(\frac{\beta}{3} \Bigr) F_{\infty} \Bigl( 
\frac{\beta}{3} \Bigr),
\end{equation}
which reveals the self-similarity with respect to the scale 
transformation
 by the factor $3$. The maximum of $F(\beta)$
occurs at $\beta = 0$, $F_N (0) = 1$, and the function 
oscillates very rapidly.
Figure (2a) shows $F_N ( \beta)$ in the case $2 ka = 3^{10}$, $N = 10$ for
$ 0 \leq \theta \leq \frac{\pi}{2}$, while Figure (2b) 
shows the same function
only for the region $0 \leq \theta \leq 0.04 $. Apparently they 
look similar, and this 
self-similarity becomes more striking as $N$ increases. The distribution
of zeroes looks random, which may imply that the distribution 
of the dark fringes 
in the 
interference pattern is fractal. 

\section{Generalization of Cantor Diffraction Grating}

For the purpose of generalizing the Cantor grating we first consider
a function representation of the Cantor set. Let $g(x)$ be a periodic function
of period $2a$, 
\begin{equation}
g(x)= \left\{\begin{array}{ll}
              ~1,  &  -a \leq x \leq  -\frac{a}{3},  \\
              ~0,  &  -\frac{a}{3} < x < \frac{a}{3},  \\
              -1, &  \frac{a}{3} \leq x \leq a \,, 
              \end{array}
       \right.
\end{equation}
and define a grating structure function $G_N (x)$ as
\begin{eqnarray}
G_N(x)   \equiv  g (x) g (3x) \cdots g ( 3^{N-1} x) 
= \prod_{m = 0}^{N-1} g( 3^m x),~~~~ ( - a \leq x \leq a).
\end{eqnarray}
Using this grating structure function the interference pattern through
the Cantor grating can be represented as
\begin{eqnarray}
A_N (\beta) &=& \int^{a}_{-a} dx e^{ikx \sin \theta} G_N (x)
\nonumber\\
&=& \Bigl(\frac{2}{3} \Bigr)^N 2a S_N (\theta) F_N (\beta),
\end{eqnarray}
where $ \beta = 2 ka \sin \theta$. In this way we construct a 
transformation between
the grating function $G_N (x)$ and the interference pattern 
function $F_N ( \beta)$
modulo the single-slit factor $S_{N} (\theta) $:
\begin{eqnarray}
G_N (x) = \prod_{m = 0}^{N-1} g( 3^m x) \leftrightarrow
F_N (\beta) = \prod_{m = 1}^{N} 
\cos \Bigl( \frac{\beta}{3^m } \Bigr) 
= \prod_{m = 0}^{N-1} \cos \Bigl(\frac{3^m \beta}{3^N} \Bigr),
\end{eqnarray}
which may be called a diffraction transformation between $G_N (x)$ and
$F_N (\beta)$. Their similarity in the product form of 
periodic functions and scaling
property are obvious. To obtain the interference pattern 
we simply replace the generating function $g(x)$ by the periodic
function $\cos \Bigl( \frac{\beta}{3^N } \Bigr)$. 

Let us now consider the possible generalization of the fractal structure.
One immediate generalization is to take the generating function $g(x)$ 
different from the simplest Cantor case. 
For example, we consider a periodic function
$h(x)$ of period $2a$. As shown in the Figure 3, the region $[-a, a]$
is divided into $2 H + 1$ 
pieces of equal length with 
$H$ being a positive integer. Its value is zero in the dark region
and one in the open region, 
where the dark and open regions alternate.
\begin{equation}
%\[
h(x)= \left\{ \begin{array}{ll}
             1,  & - a + \frac{4ma}{2H +1} \leq x  
\leq -a + \frac{(4m + 2)a}{2H + 1}, \\
             0,  & - a + \frac{(4m + 2)a}{2H +1} < x < 
-a + \frac{4(m+1)a}{2H + 1}, ~~~ 
             (m = 0, \pm 1, \pm 2, \cdots).
             \end{array}
      \right.
%\]
\end{equation}
With $H=1$, one finds that $h(x)$ reduces to $g(x)$ of the 
previous section.
From this periodic function we construct a grating function
\begin{equation}
G_N (x) = \prod_{m= 0}^{N-1} h \Bigl((2H+1)^m x \Bigr),
~~ - a \leq x \leq a,
\end{equation}  
which, in the limit $N \rightarrow \infty$, becomes a fractal 
set with the fractal
dimension,
\begin{equation}
D = \frac{\ln (H +1)}{ \ln (2 H +1)}.
\end{equation}
The Fraunhofer diffraction amplitude through this grating is
\begin{equation}
A_N  = \frac{2 \sin \Bigl(\frac{ka \sin \theta}
{ (2H +1)^N} \Bigr)}{ k \sin \theta}
\sum_{j = 1}^{(H+1)^N} e^{ ik x_j \sin \theta},
\end{equation}
where $x_j$ are the midpoints of the open slits. Defining
\begin{equation}
F_N = \sum_{j = 1}^{(H+1)^N} e^{ ik x_j\sin \theta},
\end{equation}
we find the recursive relation
\begin{equation}
F_N (\beta)  = \frac{\sin  \Bigl(\frac{(H+1) \beta}
{  (2H +1)^N} \Bigr) }{\sin  \Bigl(\frac{\beta}{ (2H 
+1)^N}\Bigr)} F_{N-1} (\beta), 
\end{equation}
where $\beta = 2 ka \sin \theta$. Proceeding as in the previous 
section, we finally  get
\begin{equation}
F_N (\beta) = \prod_{m = 0}^{ N-1}\frac{\sin \Bigl(\frac{(H+1) 
\beta}{ (2H +1)^m}\Bigr)}{\sin \Bigl(\frac{\beta}{ (2H +1)^m} \Bigr)}. 
\end{equation}
In the simplest case $ H = 1$, we confirm the previous result
\begin{equation}
F_N (\beta) = \prod_{m = 1}^{ N}\frac{\sin \Bigl(\frac{2 
\beta}{ 3^m}\Bigr)}{\sin \Bigl(\frac{\beta}{ 3^m}\Bigr)}
= \prod_{m =1}^{N} 2 \cos \Bigl( \frac{\beta}{3^m} \Bigr) . 
\end{equation}
The diffraction transformation then becomes
\begin{equation}
G_N (x) = \prod_{m = 0}^{N-1} h \Bigl( (2 H +1)^m x \Bigr) 
\leftrightarrow
F_N (\beta) = \prod_{m = 1}^{N} \frac{\sin \Bigl( \frac{ (H +1) \beta}{
(2 H +1)^m } \Bigr)}{\sin \Bigl( \frac{\beta}{ (2 H +1)^m} \Bigr)}.
\end{equation}
We note that the factor function in $F_N$ is the familiar 
Fraunhofer diffraction
pattern of a regular grating with $(H+1)$-slits whose 
width are $\frac{2a}{(2H+1)^m}$.
This clearly exhibits the self-similarity with respect to the 
scale transformation
by a factor $2H+1$.

\section{Discussion}

We have considered the Fraunhofer interference pattern 
obtained from  diffraction gratings  whose slits are 
fractally distributed.  The  simplest Cantor   set type grating   
and its  generalization were 
explicitly constructed,  and their  interference patterns  
were  studied. The  grating structure 
function, $G_N   (x) =   \prod_{m =  0}^{N-1}  g  
\Bigl( (2H+1)^m   x \Bigr)$,   gives the 
interference    pattern    
function,    $F_N   (\beta)    =    \prod_{m    =    0}^{N-1}    f 
\Bigl(\frac{\beta}{(2H+1)^m} \Bigr)$, by the  
diffraction transformations, both  of which show 
the repeated product structure, and have the self similarity with 
respect to the scaling 
by a factor of $2H+1$. However, the fractal dimension of the 
function of interference pattern $F_N(\beta)$ is not fully 
understood.

For general gratings with fractal distribution, 
analytic results may not be easily obtained. 
%contrasted to the 
%case we have discussed. 
For example, if one  tries 
the triadic Cantor generator whose segments are of unequal lengths, 
the  computations are no 
longer straightforward because the factorization is not possible. 
In this case numerical calculations 
would be interesting.  If we  consider a  generator function  
$ g(x)$  of general form,  it is 
not easy  to picturize  the product  function $G_N  (x)$ nor  
to evaluate  the diffraction 
pattern function $F_N (\beta)$. 
%One important exception  is the Gaussian form $g(x) = e^{  - 
%\alpha x^2}$ whose scale transformation by any scale 
%gives the Gaussian again with different 
%width. The diffraction pattern is also of the Gaussian form. 
%In this  sense Gaussian diffraction 
%grating is  covariant  for any   scale transformation,  
%and self-adjoint   with respect  to the 
%diffraction transformation.

Another  direction of generalization is to consider 
higher dimensional fractal diffraction 
gratings which we intend to study in future.

\section*{acknowledgements}

This work was supported in parts by the Basic Science Research 
Institute Program,
Ministry of Education under Projects Nos. BSRI-97-2418, BSRI-97-2425, 
and BSRI-97-2427,
and by the Center for Theoretical Physics, Seoul National University.
One of us (K. Soh) thanks Ms. Myung-Hwa Lee for helping him 
in numerical calculations.

\section{figure caption}

FIG. 1.
Cantor Diffraction Grating for $N=3$. The width of each open 
slit is $\frac{2a}{27}$.
$x_j, x_j^-, x_j^+$ are the middle, beginning, and 
end points, respectively, of
the $j$-th open slit.

FIG. 2a.
The amplitude factor $F_N (x)$ with $x=2ka\sin\theta$  
for $N = 10,\,\, 2 ka = 3^{10}$, 
and $0 \leq \theta \leq
\frac{\pi}{2}$.

FIG. 2b.
The same amplitude factor $F_N (x)$, 
rescaled to the range 
$0 \leq \theta \leq
0.04$. The self similarity of the function would be more 
striking if $N$ becomes 
larger.

FIG. 3.
The Cantor-like generator of $(2H+1)$-slit. It is a 
periodic 
function of period $2a$.
The region $[-a, a]$ is divided into $2H+1$ pieces of 
equal length. 
The $h(x)$ takes
values either zero or  one alternatively.

\end{document}